\documentclass[11pt,twoside]{article}

\usepackage{asp2014}

\aspSuppressVolSlug
\resetcounters

\begin{document}

\title{Cosmological Simulations in Exascale Era}

\author{D. Goz,$^1$ L. Tornatore,$^1$ G. Taffoni,$^1$ and G. Murante$^1$}
\affil{$^1$INAF - Osservatorio Astronomico di Trieste - via Tiepolo 11, 34131 Trieste Italy; \email{goz@oats.inaf.it}}

\paperauthor{D. Goz}{goz@oats.inaf.it}{orcid.org/0000-0001-9808-2283}{INAF}{Osservatorio Astronomico di Trieste}{Trieste}{Trieste}{34131}{Italy}
\paperauthor{L. Tornatore}{tornatore@oats.inaf.it}{orcid.org/0000-0003-1751-0130}{INAF}{Osservatorio Astronomico di Trieste}{Trieste}{Trieste}{34131}{Italy}
\paperauthor{G. Taffoni}{taffoni@oats.inaf.it}{orcid.org/0000-0002-4211-6816}{INAF}{Osservatorio Astronomico di Trieste}{Trieste}{Trieste}{34131}{Italy}
\paperauthor{G. Murante}{murante@oats.inaf.it}{orcid.org/0000-0002-5155-130X}{INAF}{Osservatorio Astronomico di Trieste}{Trieste}{Trieste}{34131}{Italy}

\begin{abstract}
The architecture of Exascale computing facilities, which involves millions of heterogeneous processing units, 
will deeply impact on scientific applications. Future astrophysical HPC applications must be designed to make 
such computing systems exploitable.
The ExaNeSt H2020 EU-funded project aims to design and develop an exascale ready prototype based on low-energy-consumption 
ARM64 cores and FPGA accelerators. We participate to the design of the platform and to the validation 
of the prototype with cosmological $N$-body and hydrodynamical codes suited to perform large-scale, 
high-resolution numerical simulations of cosmic structures formation and evolution.  
We discuss our activities on astrophysical applications to take advantage of the underlying architecture.

\end{abstract}

\section{Introduction}
The ExaNeSt H2020 \citep{ExaNeSt} project aims at the design and development of an exascale ready supercomputer 
with a low energy consumption profile but able to support the most demanding scientific and technical applications. 
The project will produce a prototype based on low-consumption ARM64 processors, FPGA accelerators and 
low-latency interconnections implementing a co-design approach where scientific applications requirements 
are driving the hardware design \citep{I12-1_adassxxvi}. 
Node level heterogeneous architectures, compared to traditional CPUs, offer high peak performance and 
are energy and possibly cost efficient. Computing power in existing peta-scale machines is already mainly issued by accelerators (basically GPUs). This will be exacerbated even further on future exa-scale platforms 
that will involve millions of specialized parallel computing units.
However, the majority of current available astrophysical codes for HPC have been designed and 
developed with a substantially different paradigm in mind (i.e. limited number of cores, distributed memory, 
absence of accelerators), and following the "procedural" nature of the codes. 
Instead, programmers must re-design and re-engineer their codes in order to exploit the exa-scale heterogeneous architecture, 
based on different devices and likely with complex memory hierarchies.

We participate to the design of the platform and to the validation of the prototype with 
{\sc{HiGPUs}} \citep{HiGPUs1}, a direct $N$-body code to simulate stellar cluster dynamics and close encounters, 
{\sc{PINOCCHIO}} \citep{Monaco_2002}, a code aimed to generate catalogues of dark matter halos and their merger histories, 
and {\sc{GADGET-3}}, evolution of the public code {\sc{GADGET-2}} \citep{Springel_2005}, a state-of-the-art $N$-body and hydrodynamical code for large-scale, 
high-resolution numerical simulations of cosmic structure formation and evolution.

Our ensemble of applications offers the opportunity to fully exploit the co-design concept. 
In fact, on one hand we have the chance to conceive, develop and test the new architecture, 
and on the other hand core algorithms of the aforementioned applications are going to be improved in such a way to 
increasingly fit to the exa-scale target platform. 

\section{Porting activity on FPGA}
"Unconventional" FPGA devices in comparison to both CPUs and GPUs are more power-efficient 
(i.e. higher throughput per watt) for different class of applications. 
Unlike both CPUs and GPUs, FPGAs do not have any fixed architecture. 
On the contrary, they provide fine-grain grid of functional units, such as DSP and memory blocks, 
which can be interconnected to make any desired circuit. 
The main drawback in the usage of FPGAs, however, is the complexity of programming them, 
which requires low level coding in hardware description. 
This severe limitation can be mitigated by a technique called high-level synthesis (HLS) 
that allows the conversion of an algorithm description in high level languages (e.g. C/C++ or OpenCL) into a digital circuit.

{\sc{HiGPUs}} is a direct $N$-body code written in C/C++ language and it is parallelized using MPI 
and OpenCL in order to exploit GPU clusters. 
The code relies on the Hermite 6th order time integrator scheme that ensures high accuracy 
for positions and velocities that are updated every time step. 
Double-precision (DP) in inter-particles distance and acceleration is mandatory in order 
to minimize the round-off errors, although some intermediate calculations may 
be executed in single-precision (SP) to speed up the time to solution.

To investigate the issues encountered when implementing and optimizing 
an OpenCL-based application on FPGA, we decided to focus our attention only on the most 
computationally demanding {\sc{HiGPUs}}'s kernel ($O(N^2)$ computational cost, with $N$ the number of particles). 
The strategies adopted to find a possible optimizations on FPGA, balancing throughput and resource
usage, are \emph{(i) kernel vectorization} and \emph{(ii) extended-precision (EX) floating-point numbers}.
Vectorizing code can effectively improve memory bandwidth because of regular memory access, 
better coalescing of these memory accesses and reducing the number of loads (each load/store is larger).
As mentioned above, the numerical solution of the $N$-body problem would strictly require DP
operations. However, DP arithmetic is extremely resource-eager and performance-poor in FPGAs.
As an alternative, EX numeric type can represent a trade-off in FPGAs. 
An EX number provides approximately 48 bits of mantissa at single-precision exponent ranges \citep{Thall_2006}.

The functional correctness of the modified kernel has been tested on a GPGPU Nvidia Tesla M2075 for scientific computing.
\articlefiguretwo{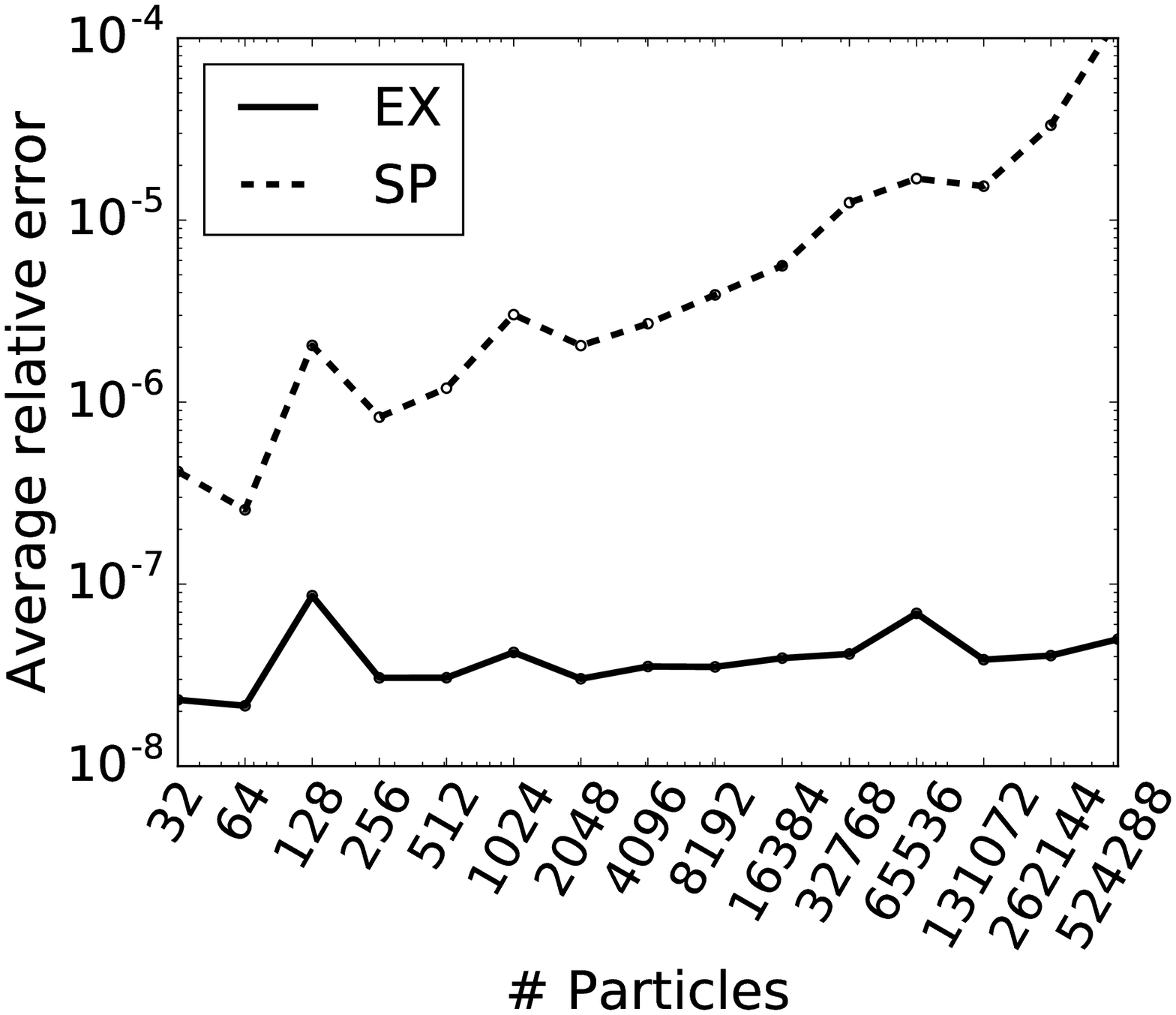}{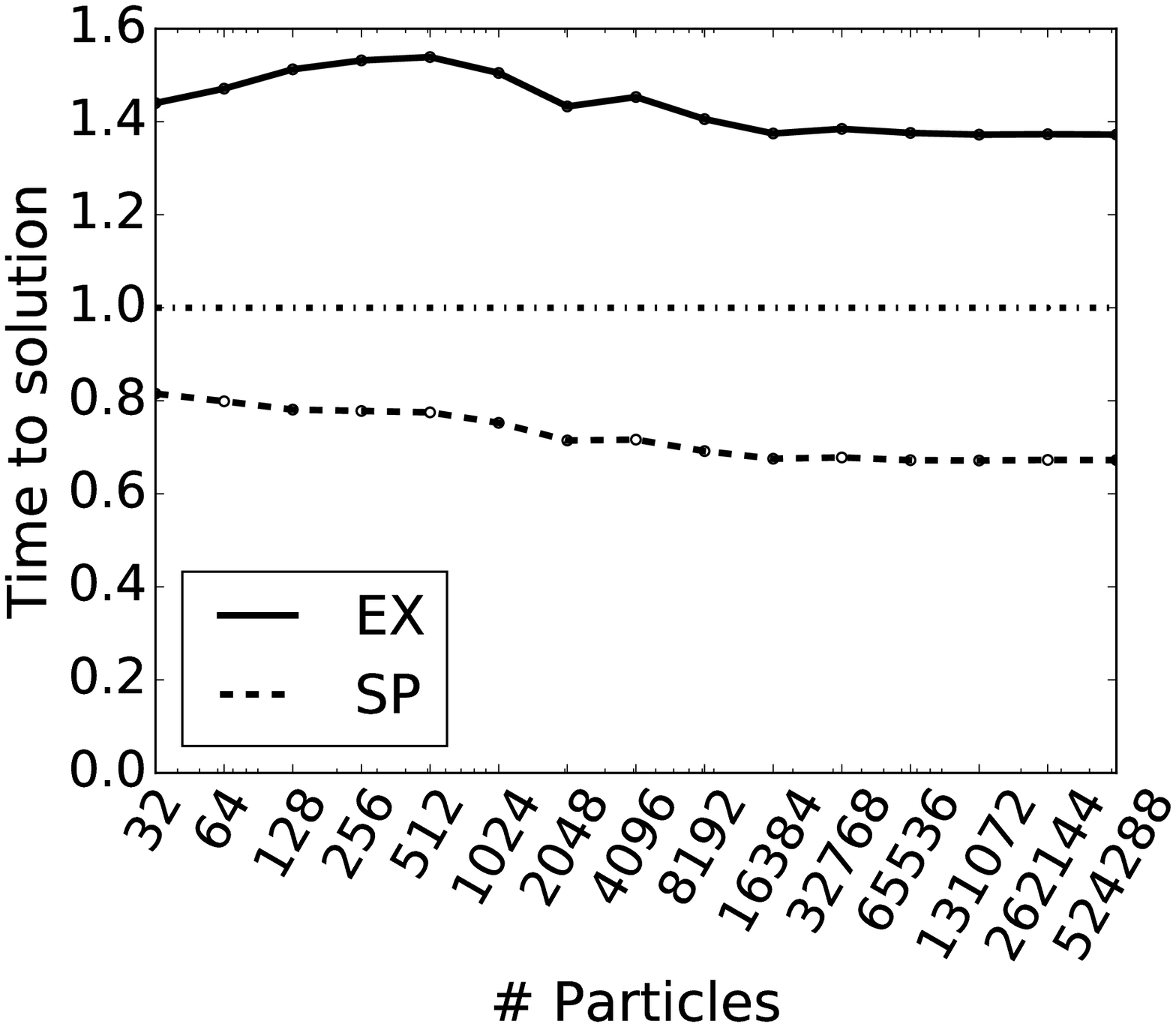}{fig1}{Test performed on GPGPU Nvidia Tesla M2075 (INFN Sezione di Roma courtesy). 
\emph{Left:} average relative error between 
EX (continuous line), SP (dashed line) arithmetics and DP arithmetic as a function of the number of particles.
 \emph{Right:} timing profile for EX (continuous line) and SP (dashed line) as a function of the number of particles. 
 Time to solution is normalized to the value obtained with DP arithmetic.}
The tests presented in Fig.\ref{fig1} suggest that EX arithmetic can be adopted for $N$-body problem ensuring 
to keep control over the accumulation of the round-off error during the simulation. However, time to solution 
reveals some overhead to handling EX arithmetic.
This approach might allow us to require only SP capability to the FPGA, saving resource usage and possibly 
not incurring in performance penalties.

\section{Code migration into Exa-scale Era}
{\sc{PINOCCHIO}} code, written in C language and parallelized with MPI, relies on the widely-employed FFTW library \citep{FFTW} required
to perform Fast Fourier Transform to make the calculations. 
The drawback of it, when used in parallel, is that it is only able to decompose the domain in one
dimension (slab decomposition).
To perform a FFT of a 3D domain with grid number $N_{0} \times N_{1} \times N_{2}$ has two major consequences: \emph{(i)} only $N_{0}$ tasks will actively perform the calculations, 
and \emph{(ii)} $N_{1} \times N_{2}$ may not fit in memory when we require exascale capabilities.
To overcome both issues listed above in the foreseen exascale architectures, we adopted the PFFT library \citep{PFFT} in place of FFTW, 
allowing both 2D and 3D decompositions (pensil decomposition), also re-designing some memory structures and memory access patterns.

Moreover, the code needs to fill a 2D plane of size $N_{0} \times N_{1}$ with pseudo-random numbers
in order to generate the initial conditions. The plane is filled following special symmetries
due to a physical-motivated recipe whose details are not discussed here.
The crucial point here is that, if one wants to generate the i.c. for a cosmological
box increasing the mass- and spatial-resolution, the ``new'' signal is added at small scales
leaving unchanged the signal at larger scales. 
Thus, this means that portions of the 2D plane filled with pseudo-random numbers would be exactly the same changing the resolution.
The code manages easily all this procedure because each MPI task retains in memory the whole plane.
However, memory-bound issues occur when $N$ is large (e.g. beyond $10^{4}$).

We performed an actual re-engineering of the i.c. generator. In order to allow each MPI task
to retain in memory only a region of the plane (i.e the plane is distributed among MPI tasks)
we were forced to figure out a completely different algorithm to built the plane.
The re-design activity on {\sc{PINOCCHIO}} code implies that now it is able to
efficiently split the workload not being memory-bound.

{\sc{GADGET-3}} code is written in C language and parallelized using an hybrid model, 
MPI+OpenMP. The code computes gravitational force using a TreePM technique, while the
hydrodynamics is solved using the Smoothed Particle Hydrodynamics approach.
Furthermore, the code contains many astrophysics modules designed to compute 
more processes, required to follow in details the formation and the evolution of cosmic structures.
{\sc{GADGET}}, as the majority of similar codes, has been designed and developed 
with the paradigm of a limited number of CPUs ($\sim 10^{3}$), each with a limited number of cores, distributed memory (few GB per CPU), almost exclusive use of MPI and little multi-threading, resulting in a ``monolithic'' workflow instead of being split-up in autonomous tasks running concurrently.
Secondly, {\sc{GADGET}} memory model is not NUMA-aware and memory locality is exploited only with some basic numerical stratagem.

The main bottleneck of simulations with a very large dynamical range (details in \citet{Murante_2015}) is load imbalance, arising when the calculations on a task depend on those performed on another task.
To exploit exascale architecture is mandatory to re-think from scratch the infrastructural operations, starting from re-designing them in a task-based and data-driven perspective.

We have developed mini-apps extracted from {\sc{GADGET-3}} code for each of the core algorithms, such as domain-decomposition, gravity and hydrodynamical solvers. We plan to design those algorithms as atomic inter-dependent tasks developing different strategies for each of the modules detailed above.

\section{Conclusions}
Astronomers will be forced to re-engineer their applications in order to exploit new Exascale computing facilities based on heterogeneous hardware (CPUs/GPUs/FPGAs). Re-design of the code must start from re-design of the memory model in order to exploit more efficiently
memory affinity and locality, and secondly it is compulsory to translate the workflow in a queue system where idling threads perform the first available task.

\acknowledgements The project has received funding by the European Union's Horizon 2020 Research and Innovation Programme under the ExaNeSt project (Grant Agreement No. 671553)

\bibliography{O2-2}  

\end{document}